\documentclass[10pt,a4paper]{article}
\usepackage{graphicx}
\usepackage{epstopdf}
\usepackage{amssymb,amsmath}\usepackage{latexsym}

\title{Rule-based Modeling and Simulation of Biochemical Systems with
Molecular Finite Automata}

\author{Jin Yang\thanks{Chinese Academy of Sciences -- Max Plank Society Partner Institute for Computational Biology, Shanghai Institutes for  Biological Sciences, Shanghai 200031, China. E-mail: jinyang2004@gmail.com}, Xin Meng$^1$, William S. Hlavacek\thanks{{Theoretical Division and Center for Nonlinear Studies, Los Alamos National Laboratory, Los Alamos, NM 87545, USA}}$^,$\thanks{Department of Biology, University of New Mexico, Albuquerque, NM 87131, USA}}

\date{}
\begin{document}

\maketitle
\begin{abstract}
We propose a theoretical formalism, molecular finite automata (MFA), to describe individual proteins as rule-based computing machines. The MFA formalism provides a framework for modeling individual protein behaviors and systems-level dynamics via construction of programmable and executable machines. Models specified within this formalism explicitly represent the context-sensitive dynamics of individual proteins driven by external inputs and represent protein-protein interactions as synchronized machine reconfigurations. Both deterministic and stochastic simulations can be applied to quantitatively compute the dynamics of MFA models. We apply the MFA formalism to model and simulate a simple example of a signal transduction system that involves a MAP kinase cascade and a scaffold protein.

\vspace{0.4in}


\end{abstract}

\section{Introduction}

In computational systems biology, studying a complex biochemical system involving a large number of interacting proteins often relies on {\it in silico} simulations to analyze and predict system behaviors~\cite{kitano2002csb}. In recent years, computational models have been increasingly used in cell signaling research and have been developed to study various pathways~\cite{kholodenko2006cell,aldridge2006physicochemical}. However,  models often fail to capture the mechanistic details of signal transduction systems~\cite{breitling2005current}. For example,  models sometimes inadequately account for the complexities of protein interactions, including interaction details at the level of protein sites and structural relationships among components of signaling proteins~\cite{hlavacek2003complexity}, particularly multisite protein modification in the context of multiprotein complexation~\cite{yang2005multisite}. Proteins in a signal-transduction system often have multiple component parts that enable the protein to interact with other molecules in a modular manner~\cite{hunter2000,Pawson2003,bhattacharyya2006domains}. Models that account for the functions of the component parts of proteins (e.g, linear motifs and protein interaction domains) are needed to better understand the dynamics of signal-transduction systems~\cite{hlavacek2009complexity,mayer2009molecular}.

Limitations of conventional modeling approaches, which rely on explicit specifications of chemical reaction networks, lie in both model construction and simulation. Conventional models are essentially specified with lists of biochemical species and their reactions. However, representing a biochemical system as a chemical reaction network is often cumbersome and unnecessary~\cite{hlavacek2006rules,danos2007rule}. Graphical rule-based modeling formalisms and associated simulation algorithms have been developed to represent biochemical systems in terms  of formal rules for biomolecular interactions~\cite{hlavacek2006rules,danos2007rule,Danos2004, priami2005beta,faeder2005rule,blinov2006graph,andrei2007graph,feret2009internal}. In graphical rule-based modeling, graphs (or the equivalents) are used to represent molecules, and graph-rewriting rules (or the equivalents) are used to represent molecular interactions. A rule represents a molecular interaction explicitly and the reactions that can arise from the interaction implicitly, and a rule can be viewed as a coarse-grained description of a class of reactions.

Two common types of protein interactions, multivalent protein binding and multisite post-translational protein modification, cause a combinatorial increase in the size of a reaction network with an increase in the number of interaction modules. It is usually difficult and error prone to manually construct a full-sized chemical reaction model. As an alternative, such conventional models can be automatically obtained using rule-based reaction generation tools, such as BioNetGen, Virtual Cell or little b~\cite{blinov2004bionetgen,moraru2008virtual,faeder2009rule,mallavarapu2009programming}, Moleculizer or Smoldyn~\cite{Lok2005NatB,andrews2010detailed}, Simmune~\cite{meier2006key}, or Stochastic Simulator Compiler (SSC)~\cite{MieszkoLis09012009}. Unfortunately, the number of reactions and biochemical species implied by rules can be enormously large (even infinite or limited only by the number of molecules in a system) for rule-based models of signal transduction systems~\cite{hlavacek2006rules}, making it inefficient to construct, simulate and analyze conventional models derived from rules.

In addition to formalisms based on graph rewriting, a number of theoretical frameworks for biomolecular interaction systems have been proposed over the past decade or so to facilitate model building and simulation. Despite the differences in their syntactical and grammatical structures, most formalisms share a common feature: molecular entities are treated as computational agents that interact with one another according to a collection of specific protocols~\cite{morton1998predicting,regev2001ras,priami2001asn,regev2002cac,fisher2007ecb}.  For example, protein interactions have been viewed as concurrent processes and have been modeled with communication protocols by process algebras, such as $\pi$-calculus~\cite{regev2001ras,priami2001asn}. Many of these formalisms have been coupled to Gillespie's stochastic simulation algorithm~\cite{priami2001asn,gillespie2007stochastic} to enable discrete-event simulation.

In engineering and computer science, complex dynamical systems with heterogeneous, modular and reactive components are frequently modeled by state machines and related formal structures. In this paper, we propose a new formalism, referred to as {\it molecular finite automata} (MFA), to model individual proteins as structured computing agents and to specify protein-protein interactions in the form of synchronized dynamics of interacting agents. The main goal is to provide an intuitive as well as programmable representation for biomolecular interaction systems. The MFA formalism is developed by incorporating and extending the classic structure of finite automata, which is a well-established formalism that has a wide application range and for which many sophisticated software and hardware tools are available. As we will see, an agent within the MFA framework explicitly represents a protein's activity (or state) induced by external inputs, and a protein interaction is specified as a joint transformation of the states of multiple MFA agents. At the systems level, a collection of {\em reaction rules} is used to describe interactions among the MFA agents in a system.

\begin{figure}[t]
\centering
\includegraphics[scale=0.5]{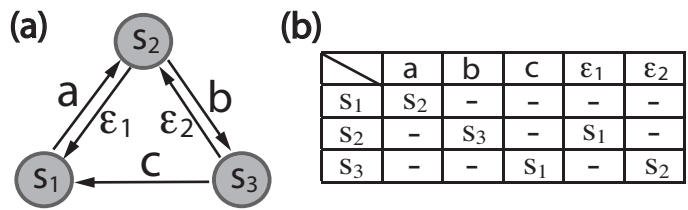}
\caption{\label{fig:fsm} An example finite automaton. (a) State transition diagram for a three-state finite automaton. A circle denotes a state. An arrow denotes a state transition. A letter next to an arrow denotes an input. (b) A state transition table for the finite automaton in panel (a). The leftmost column indicates possible current states. The topmost row indicates inputs. A table entry indicates a target state given a current state and an input. The symbol `--' indicates ``not applicable."}
\end{figure}

We also report simulation methods that can compute the dynamics of a system modeled within the MFA framework. Using the example of a MAP kinase cascade, we demonstrate how to apply the MFA formalism to model and simulate a cell signaling system. 

\section{Formal model}
In the first part of this section, we introduce a representational framework using MFAs to describe the building blocks, particularly proteins, of biomolecular interaction systems. In the second part, we show how to apply the MFA representations to construct quantitative models for cell signaling systems and how to compute with these models. 

\subsection{Molecular entities --- molecular finite automata}
The notion of finite automata, or finite state machines, is well-established in theoretical computer science and has been applied to model the dynamics of diverse discrete systems (i.e., systems with finite numbers of states)~\cite{Hop-Auto2000}. Finite automata were traditionally developed to construct parsers and compilers, conduct formal verifications and mathematical proofs, and design and test software~\cite{Hop-Auto2000,sipser2005introduction,LeeD:Priamt}. Because of their simple, adaptive and intuitive structure, finite automata have been applied in a number of areas, including engineering systems design~\cite{borger2003abstract}, computational linguistics~\cite{mohri1997finite} and communication protocols~\cite{brand1983communicating}. Interacting state machines have also been used in the field of computational biology to visualize and model cellular level interactions~\cite{fisher2007ecb}.

Our goal in this paper is to propose a formalism based on an extended structure of finite automata that is suited for modeling biomolecular intreactions at the submolecular level with consideration of site-specific details. Below, we give a formal definition of a purely reactive finite automaton that will be extended for the later description of biomolecules.

\noindent {\bf Definition 1 (finite automaton).} {\em A finite automaton is a tuple $D=(S, X, \delta, s_0)$, where $S$ and $X$ are finite sets of states and inputs, respectively. The function $\delta$ is a transition function that maps the current state along with an input in $X$ into a target state, $\delta: S\times X \to S$. The symbol $s_0$ denotes a start state.}

The automaton of Definition~1 is a so-called ``deterministic'' finite automaton (DFA). In a DFA, given an input, a state transition is non-ambiguous, and a DFA can only reside in one state at any given time. In contrast, a finite automaton can be nondeterministic, in which multiple state transition paths (or more than one target state) may exist for a single input. Here, assuming responses of a protein to external signals are deterministic, we focus on DFAs as the fundamental structures for modeling proteins.

The above definition characterizes a finite automaton as a reactive model in which state transitions are induced by external events in the form of input signals. Elements in the set of states $S$ are represented using subscripted lower-case letters, $S=\{s_1,s_2,...,\}$. Throughout, inputs are represented using the lower-case alphabet, $\varepsilon$ and subscripted $\varepsilon$'s, i.e., $X=\{a,b,...,\varepsilon,\varepsilon_1, \varepsilon_2,...\}$. A special input symbol $\varepsilon$ is introduced to model a non-specific external signal or a signal from an unknown source that causes a spontaneous state transition. In a model of a signaling system, such a non-specific input can be used to model molecular events such as dissociation of two bound proteins caused by background collisions with solvent molecules or protein modifications catalyzed by unknown enzymes. A finite automaton can be visually represented by a {\it state transition diagram} (Fig.~\ref{fig:fsm}(a)), a directed graph in which a node denotes a state and an edge denotes an input-induced transition. Equivalently, a finite automaton can be specified by a machine-readable {\it state transition table} (Fig.~\ref{fig:fsm}(b)).

The dynamics of many reactive systems can be represented using the DFA structure of Definition~1. However, this structure is inefficient for describing signaling proteins. To extend the DFA structure to model a protein, we first look at the correspondence between properties of finite automata and protein functions. A classic finite automaton models a memoryless process, wherein a state transition depends only on the current state and an input. In contrast, protein interactions mostly happen under certain molecular contexts. To see the importance of molecular context, we consider allosteric regulation and protein complexation. Allosteric regulation of a protein or enzyme is a common mechanism in biochemistry. Protein activity in one domain is changed (either activated or inhibited) by binding or unbinding of an effector molecule at another site. The formation of heterogeneous and transient multiprotein complexes is one of the essential functions of protein-protein interactions in signal transduction. Context-sensitive interactions such as co-localization of an enzyme and one of its substrates control both the strength and specificity of molecular signaling. These features of protein interactions require an extension of the DFA structure beyond the representation of information only in terms of a finite number of states.

\begin{table}[t]
 \centering
\caption{Operators and symbols used in MFA structures\label{tab:op}}
\begin{tabular}{cl}\hline
 \bf Operator/symbol & \bf Definition \\ \hline\hline
$x := a$ & Assignment of a value $a$ to a variable $x$\\
$x=a$ & Comparison between $x$ and $a$\\ 
$/ a$& Delimiter that precedes an operation $a$\\
$\backslash a$ & Delimiter that precedes a predicate $a$ \\ 
$a.b$ & Component operator: $b$ is a member of $a$ \\
$A$--$B$ & Bond association between $A$ and $B$ \\
$x\to A$ & Mapping input $x$ to machine $A$ \\
\hline
\end{tabular}
\end{table}

To capture the contextual sensitivity of protein interactions, internal variables are introduced to record contextual information such as information about  binding partners or other local molecular information. An extension of Definition~1 should also include functions that will read and modify the machine variables. Along with state transitions, these machine operations update the configuration of a protein. Based on these considerations, we define an enhanced automaton structure, an ``extended finite automaton" (EFA) to amend the classic finite automata structure.

\noindent {\bf Definition 2 (extended finite automaton).} {\em An extended finite automaton is a tuple $E=(S, X, \delta, s_0, \mathbf{v})$, where $S$ and $X$ are finite sets of states and inputs, respectively. The transition function, $\delta: S\times X \backslash P(\mathbf{v}) \to S / A(\mathbf{v})$, maps the current state along with an input in $X$ into a target state upon evaluation of a predicate function $P(\mathbf{v})$, and performs an operation $A(\mathbf{v})$ on the variable structure $\mathbf{v}$ along with the state transition. The symbol $s_0$ denotes a start state.}

The meanings of operators (e.g., $\backslash$ and $/$) used in the above definition are given in Table~\ref{tab:op}. Our definition of EFA is close to the convention of an extended finite state machine~\cite{LeeD:Priamt}, which also involves operations on internal variables. Suppose that an EFA $E$ is in state $s$. Upon receiving an input $x$, $E$ undergoes a transition $\delta=(s,q,x,P(\mathbf{v}),A(\mathbf{v}))$, where $s$ and $q$ are the source and target states, respectively. If the predicate $P(\mathbf{v})$ is true (e.g., an evaluation of variables in $\mathbf{v}$ indicates that the transition is legitimate), $E$ moves to the target state $q$ and performs an operation $A(\mathbf{v})$ on the variable structure $\mathbf{v}$.

Ultimately, another important and ubiquitous feature of cell signaling, site-specific interactions, must be incorporated to reflect the modularity of protein interactions. Many signaling proteins possess multiple functional motifs, domains and sites, which serve as modules for combinatoric protein organizations that can potentially generate diverse signaling patterns. A realistic protein automaton should express dynamics at the level of protein sites. To this end, we arrive at the definition of ``molecular finite automaton" (MFA), which models the discrete dynamics of a multidomain biomolecule. The relationship between MFA and EFA is as follows: (1) an MFA contains one or multiple EFAs and (2) each EFA in an MFA operates on a common variable structure that is shared by all EFAs. Table~\ref{tab:p2a} summarizes a conceptual mapping between protein functions and the structure of an MFA.

\noindent {\bf Definition 3 (molecular finite automaton).} {\em A molecular finite automaton is a tuple $M=(E_1, E_2, ..., E_n, \mathbf{v})$, which is composed of $n$ component EFAs and a shared variable structure $\mathbf{v}$. The transition function for $E_i$ is $\delta_i: S_i\times X_i \backslash P_i(\mathbf{v}) \to S_i /A_i(\mathbf{v})$.}

Table~\ref{tab:op} lists a set of operators and symbols that we will use to describe MFAs in state transition diagrams and state transition tables. In essence, the MFA structure encapsulates multiple finite automata and allows for a hierarchical description of the component substructures of proteins. Internal variables and predicates help to compress the state space and make an MFA more accessible to intuitive understanding. Without using internal variables and predicate functions, one can still build an MFA by expanding the state space assuming that variables store information of finite size. However, such an approach may result in a state expansion that might become intractable for a complex system.

\begin{table}[t]
\centering
\caption{\label{tab:p2a} Molecular finite automaton and protein function} 
\begin{tabular}{ll}\hline \bf MFA component & \bf Protein \\\hline\hline
State & Conformation\\
State transition & Conformation change\\
Input & Biochemical interaction\\
Variable and predicate & Molecular context \\
Component machine & Domain or site \\\hline
\end{tabular}
\end{table}

The construction of an MFA requires knowledge and/or a hypothesis about the biochemistry and the component substructure of the protein one wants to model. Although some proteins have established functions in well-studied signaling pathways, biochemical mechanisms for many protein functions still await characterization. For a protein with known structure and function, the corresponding MFA must be designed to faithfully reproduce the reactive dynamics of the protein. For a poorly characterized protein, building an MFA, as in building any model, provides an opportunity to generate testable hypotheses.

\begin{figure}[t]
\centering
\includegraphics[scale=0.5]{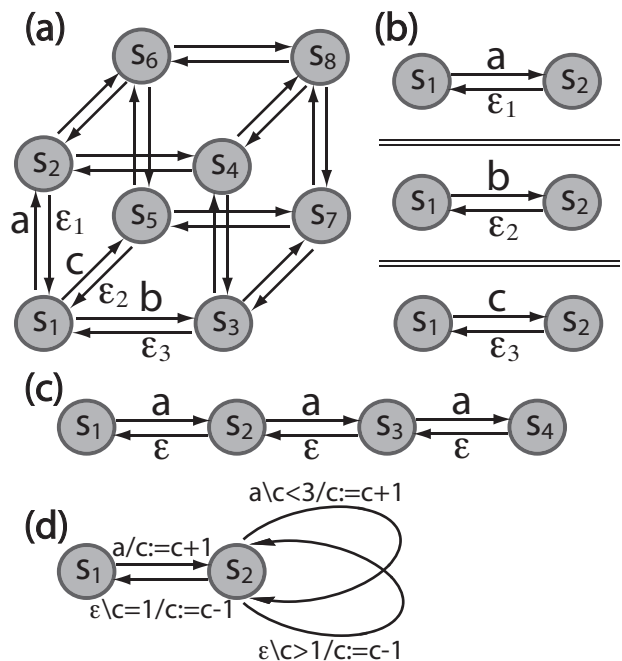}
\caption{\label{fig:idb} A protein with three binding domains modeled by different MFA structures. (a) An MFA that uses a single eight-state EFA to model the overall state transitions. Unlabeled transitions are induced by the same inputs as those identified for parallel transitions. (b) An MFA that models the protein with three independent internal finite automata, each of which interacts with distinct binding partners implied by input symbols $\{a,b,c\}$. Spontaneous inputs ($\varepsilon_1$, $\varepsilon_2$ and $\varepsilon_3$) are distinguished for non-identical individual EFAs. (c) A four-state MFA that models the protein as having three independent and identical binding sites. Machine variables that register binding partners are not shown. (d) A two-state MFA that can replace the model in (c). The two states $s_1$ and $s_2$ represent ``free" and ``bound", and the variable $c$ counts the number of bound sites.}
\end{figure}

As a design issue, MFA models can be constructed with a great deal of flexibility. Equivalent MFAs may differ in the number of states and the topology of state transition diagrams. A protein with multiple sites can be modeled by an MFA that has separate finite automata, each of which describes the dynamics of a domain. Equivalently, instead of using one automaton to model one protein site, the protein can be modeled by an MFA that consists of a single finite automaton that describes the combined behavior of all sites. For example, if a biomolecule has three independent domains that interact with different binding partners, it can be modeled as one eight-state finite automaton plus a variable structure (Fig.~\ref{fig:idb}(a)), where state $s_1$ indicates that the molecule is in a free form with no binding partners and state $s_8$ indicates that all sites are occupied. Alternatively, the biomolecule can be modeled with three two-state (a free state and a bound state) EFAs with each EFA describing an individual binding domain (Fig.~\ref{fig:idb}(b)). For the case of three identical and non-cooperative binding domains, it may be preferable to model the protein with a four-state finite automaton with the state space $S=\{s_1: \text{free}, s_2: \text{singly-bound}, s_3: \text{doubly-bound}, s_4: \text{triply-bound}\}$ for a parsimonious structure in terms of the number of states (Fig.~\ref{fig:idb}(c)). This four-state MFA can be further compressed to a two-state model as shown in Fig.~\ref{fig:idb}(d), where $s_1$ denotes the unoccupied state and $s_1$ denotes the protein is occupied at least on one of its three sites. In this model, the information about how many sites are bound is resolved by a variable $c$ serving as a counter.

The state space of an MFA, $S_M$, is a subset of the product of the state spaces of component EFAs, i.e., $S_M \subseteq {S_1\times S_2 \times ...\times S_n}$, where the two sides achieve equality when all component EFAs are independent. The input set of an MFA is a union $X_M=\bigcup_{i=1}^nX_i$. For an input $x\in X_M$, the transition function $\delta_i$ is chosen if $x$ only belongs to $X_i$. A transition function is chosen arbitrarily if $x$ belongs to input sets of multiple component EFAs. For example, if $x\in X_i\cap X_j$, either $\delta_i$ from $E_i$ or $\delta_j$ from $E_j$ can be equivalently chosen to react to the input $x$. This scenario of relaying inputs corresponds to the case where a protein has multiple domains that interact with identical partners.

\begin{figure}[t]
\centering 
\includegraphics[scale=0.5]{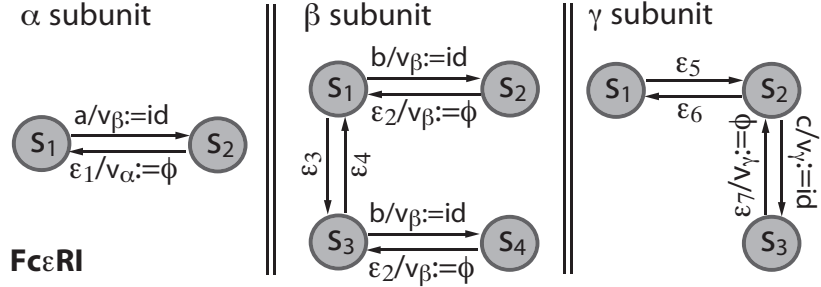}
\caption{\label{fig:fceri} State transition diagram for the MFA of a receptor Fc$\varepsilon$RI with three component EFAs. $\alpha$ subunit: unbound ($s_1$), bound ($s_2$); $\beta$ subunit: unbound and unphosphorylated ($s_1$), bound and unphosphorylated ($s_2$), unbound and phosphorylated ($s_3$), and bound and phosphorylated ($s_4$); $\gamma$ subunit: unbound  and unphosphorylated ($s_1$), unbound and phosphorylated ($s_2$), and bound and phosphorylated ($s_3$). Internal variables $v_\alpha$, $v_\beta$ and $v_\gamma$ record binding partners of the $\alpha$, $\beta$ and $\gamma$ subunits, respectively. Inputs and operations (if any) are labeled together on the transition edges, separated by a delimiter symbol $/$ (cf. Table~\ref{tab:op}). For example, $\varepsilon_1/v_\alpha:=\phi$ indicates that the MFA receives a non-specific input $\varepsilon_1$ and then sets the variable $v_\alpha$ to the null value $\phi$.}
\end{figure}

To present a biological example, Fig.~\ref{fig:fceri} shows the state-transition diagram of an MFA model of the high-affinity IgE receptor, Fc$\epsilon$RI,  in the model of Goldstein et al.~\cite{goldstein2002modeling} and Faeder et al.~\cite{FaederJamesR.:InveeF}. The state transition table for the MFA representation of Fc$\epsilon$RI is shown in Table~\ref{tab:betaitam}. The receptor molecule Fc$\epsilon$RI has three functional domains: (1) an extracellular $\alpha$ subunit responsible for binding its ligand, IgE (in fact, an IgE dimer is considered in the models in Refs.~\cite{goldstein2002modeling,FaederJamesR.:InveeF}); (2) an intracellular $\beta$ subunit that constitutively binds to Src-family protein tyrosine kinase Lyn when it is unphosphorylated and recruits Lyn with higher affinity upon phosphorylation; and (3) an intracellular $\gamma$ subunit that recruits another protein tyrosine kinase Syk upon phosphorylation. The machine for Fc$\epsilon$RI has three variables, $\mathbf{v}=(v_\alpha, v_\beta, v_\gamma)$, which record the labels (id's) of binding partners for each of the corresponding domains. We note that recording binding partners using internal variables is equivalent to constructing an adjacency list to store an undirected graph. Tracking protein connectivity by such means allows protein complexes to be represented implicitly. The connectivity of proteins within a complex can be retrieved by a graph traversal. In the Fc$\epsilon$RI pathway, some protein state transitions only happen in specific molecular contexts. For example, crosslinking of two receptors by an IgE dimer initiates signaling. On the cytoplasmic side of a crosslinked receptor dimer, a $\beta$ subunit-associated Lyn can transphosphorylate the $\beta$ subunit of the other receptor to initiate an intracellular signaling cascade~\cite{FaederJamesR.:InveeF}. To incorporate such non-local contextual information into the MFA-based pathway model of a signaling system, one needs to specify {\it reaction rules}. In the following section, we introduce a formal definition of reaction rules, which are used to describe interactions between proteins modeled by MFAs.

\subsection{Molecular interactions --- reaction rules}

An MFA is essentially a discrete state model that characterizes a protein as a reactive agent with state transition protocols. Since specification of an MFA structure does not require consideration of the modeled protein within the larger context of a signaling system, explicit rules are needed to connect individual types of MFAs as parts of an interacting system.

\begin{table}[t]
\centering
\caption{State transition table for the MFA of Fc$\epsilon$RI. \label{tab:betaitam}}
\begin{tabular}{|c|c|c|}\hline
Fc$\epsilon$RI.$\alpha$ & a & $\varepsilon_1$ \\ \hline
$s_1$ & $s_2$ / $v_\alpha := {\rm id}$ & -- \\\hline
$s_2$ & -- & $s_1$ / $v_\alpha:= \phi$ \\\hline
\end{tabular}

\begin{tabular}{|c|c|c|c|c|}\hline
Fc$\epsilon$RI.$\beta$ & b & $\varepsilon_2$ & $\varepsilon_3$ & $\varepsilon_4$ \\ \hline
$s_1$ & $s_2$/$v_\beta:={\rm id}$ & -- & $s_3$  & -- \\\hline
$s_2$ & -- & $s_1$/$v_\beta := \phi$  & -- & -- \\\hline
$s_3$ & $s_4$/$v_\beta := {\rm id}$ & -- & -- & $s_1$ \\\hline
$s_4$ & -- & $s_3$/$v_\beta := \phi$ & -- & -- \\\hline
\end{tabular}

\begin{tabular}{|c|c|c|c|c|}\hline
Fc$\epsilon$RI.$\gamma$ & c & $\varepsilon_5$ & $\varepsilon_6$ & $\varepsilon_7$ \\ \hline
$s_1$ & -- & $s_2$  & -- & -- \\\hline
$s_2$ & $s_3$/$v_\gamma := {\rm id}$ & -- & $s_1$  & -- \\\hline
$s_3$ & -- & -- & -- & $s_2$/$v_\gamma := \phi$  \\\hline
\end{tabular}

\footnotesize\noindent $\phi$: a null symbol indicating a free site. An id is a label assigned to identify an individual MFA agent among a population of agents of one type. 
\end{table}

A protein-protein interaction system is composed of a collection of MFAs for different types of proteins in the system. To describe the interactions between these MFAs in terms of biochemical reactions, one can specify protein interactions for the MFAs by means of reaction rules. An interaction between proteins changes the states of all participating molecules. In other words, a reaction synchronizes state transitions and machine reconfigurations among participant MFAs. We can view a reaction rule for an MFA-based interaction as a specification of synchronized state transitions and operations on internal variables. We formally define a reaction rule as follows. 

\noindent {\bf Definition 4 (reaction rule).} {\em A reaction rule is an injective function $R: X\to M \backslash P$. The sets $X=\{x_1,x_2,...,x_n\}$ and $M=\{M_1,M_2,...,M_n\}$ are ordered and contain inputs and MFAs, respectively. $P$ is a predicate for the mapping.}

The mapping $X\to M$ simultaneously sends each individual input $x_i$ to machine $M_i$ and executes the machine reconfigurations if $M_i$ is responsive to $x_i$. The predicate $P$ specifies an application condition for the mapping, which usually constitutes non-local molecular contexts (or patterns). Although the number of MFAs simultaneously involved in a reaction could in principle be any finite number $n$, we focus on two types of elementary interactions: (1) unimolecular interactions that involve state transitions of one MFA and (2) bimolecular interactions that involve synchronized state transitions of two MFAs. Execution of a reaction rule changes the configurations of participant MFAs according to the protocols defined in the state transition tables of the MFAs. The above definition of a reaction rule requires that MFA agents be in states that can respond to the inputs. Together with the state transition tables for individual MFA types, reaction rules provide executable and programmable protocols to connect standalone MFAs into an interacting system. 

\begin{figure}[t]
\centering
\includegraphics[scale=0.5]{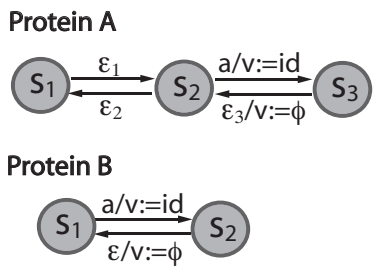}
\caption{\label{fig:react} Interactions between MFAs. An example system that is modeled by two interacting MFAs, $A$ and $B$. Automaton $A$ models a protein that can, upon phosphorylation, bind to another protein modeled by Automaton $B$. Automaton $A$ has three states: free and unphosphorylated ($s_1$), free and phosphorylated ($s_2$), and bound and phosphorylated ($s_3$). Automaton $B$ has two states: free ($s_1$) and bound ($s_2$). Internal variable $v$'s in $A$ and $B$ are used to record information about binding partners (i.e., the label of an MFA agent, or $\phi$ if free).}
\end{figure}

In the example model of Fig.~\ref{fig:react}, all interactions in the model can be specified by four reaction rules (Table~\ref{tab:mod}). Phosphorylation and dephosphorylation of automaton $A$ are approximated as unimolecular, single-step reactions, which can be defined by two rules, $R_1: \{\varepsilon_1\}\to\{A\}$ and $R_2: \{\varepsilon_2\}\to\{A\}$, respectively. In these cases, a rule is merely a local pairing of a current state and an input for an MFA. The state transition and its associated operations follow the protocol defined in the state transition table. For a bimolecular association reaction between automata $A$ and $B$, a reaction rule can be formulated as $R_3$: $\{a,a\}\to\{A,B\}$, where the MFA set and the input set are both ordered and have a one-to-one mapping. We note that the definition of a reaction rule does not specify machine states and therefore only MFAs in proper states will respond to an input. A rate law specifies a mathematical formula to calculate the kinetic rate for a reaction rule, which can be used for quantitative simulations. We note that only machines in states designated by a reaction rule are accounted for when one calculates the rate according to the rate law. For example, $R_3$ in Table~\ref{tab:mod} specifies a mass action rate law for the association reaction between protein $A$ and $B$, $r_3(t)=k_3A(\cdot)B(\cdot)$, where $A(\cdot)$ and $B(\cdot)$ represent the eligible populations of protein $A$ and $B$. Eligible machine states in an MFA can be automatically resolved by searching the state transition table for responsive states with regard to the input symbol specified in the rule. In this case, since the eligible machine states for this reaction rule are $s_2$ and $s_1$ for $A$ and $B$, respectively, the actual rate should be calculated as $r_3=k_3A(s_2)B(s_1)$. The dissociation rule, $R_4: \{\varepsilon_3,\varepsilon\}\to \{A,B\}\backslash A$--$B$, has a predicate $A$--$B$ that requires $A$ and $B$ must share a bond. The operator `--' denotes a bond association between the two machines.

In summary, a list of reaction rules assumes three roles: (1) Assigning rate laws for quantitative computation; (2) synchronizing state transitions for bimolecular reactions; and (3) making a modeling choice to decide which subsets of machine transitions are to be included in a system, in which the specification of a set of reaction rules reflects the choice of modeling assumptions and scope. For example, some state transitions may never be triggered by a given set of reaction rules even though these transitions may be possible at the machine level.

\section{Quantitative modeling}
Reaction rules are essentially specifications of coupled chemical processes that can be taken to change the configuration of a system in time. A set of rules can be translated into quantitative models if the rules can be associated with rates via rate laws. A straightforward way to  translate a rule-based model into a quantitative model is to automatically generate a conventional chemical reaction network by evaluating reaction rules using a rewriting approach~\cite{blinov2004bionetgen,faeder2005rule}. However, a far more efficient approach is to use reaction rules to directly perform a simulation. Below, we describe how to construct and simulate models specified in terms of MFA structures, for either deterministic or stochastic simulation.

\begin{table*}[t]
\begin{center}
\caption{Formal reaction rules for the model of Fig.~\ref{fig:react}\label{tab:mod}}
\footnotesize
\begin{tabular}{llll}\hline \bf Rule description & \bf Formal specification & \bf Rate law\\ \hline\hline
$R_1:$ Phosphorylation of $A$ & $\{\varepsilon_1\}\to\{A\}$ & $r_1(t)=k_1A(\cdot)$ \\
$R_2:$ Dephosphorylation of $A$ & $\{\varepsilon_2\}\to\{A\}$ & $r_2(t)=k_2A(\cdot)$ \\
$R_3:$ $A$ and $B$ association & $\{a,a\}\to\{A,B\}$ & $r_3(t)=k_3A(\cdot)B(\cdot)$\\
$R_4:$ $A$ and $B$ dissociation & $\{\varepsilon_3,\varepsilon\}\to \{A,B\}\backslash A$--$B$ & $r_4(t)=k_4A(\cdot)$, or $k_4B(\cdot)$\\
\hline
\end{tabular}

{\footnotesize \noindent \ \ Rules are shown as one-to-one mappings between ordered sets
(e.g., $R_4:$ $\{\varepsilon_3,\varepsilon\}\to \{A,B\}$ 

\ \ indicates that $\epsilon_3$ is an input for $A$ and $\epsilon$ is an input for $B$.) $A$--$B$ denotes that machines $A$ and $B$ have a bond association.}
\end{center}
\end{table*}

\subsection{Deterministic simulation}

A biochemical reaction system is conventionally modeled using coupled ordinary differential equations (ODEs) that describe the temporal evolution of all chemical species in the system. Here, we demonstrate that one can use a set of ODEs to instead describe the population dynamics of MFA states. In fact, an MFA state (or a combination of states) corresponds to an ensemble of chemical species, which often manifests as an experimental observable, such as free protein concentration or protein phosphorylation level. This idea is related to the concept of model reduction~\cite{feret2009internal,borisov2005signaling,conzelmann2006domain,borisov2006trading,borisov2008domain}, in which a reduced system of ODEs is formulated to describe the dynamics of a set of observable quantities instead of concentrations of an exhaustive set of chemical species.

In a most general form, we can write the following ODE to model the population level of the agents of MFA $M$ in state $s$, denoted as $M(s,t)$:
\begin{equation}
\frac{dM(s,t)}{dt}=r_{\rm in}(t)-r_{\rm out}(t) \ ,
\end{equation}
where $r_{\rm in}$ ($r_{\rm out}$) is the rate of population influx (outflux), consistent with the rate laws associated with the transition rules related to state $s$. For a single MFA agent, the above equation describes the time rate of change of the probability to find the machine in state $s$. 

Considering the simple example model illustrated in Fig.~\ref{fig:react}, we assume the law of mass action for protein association reactions and single-step protein phosphorylation and dephosphorylation reactions. The following differential equation can be used to describe the concentration of protein $A$ in the machine state $s_2$, $A(s_2)$:
\begin{eqnarray}
\frac{dA(s_2,t)}{dt} & = & \underbrace{k_1A(s_1,t)+k_4A(s_3,t)}_{r_{\rm in}}-\underbrace{A(s_2,t)(k_2+k_3B(s_1,t))}_{r_{\rm out}} \notag \\
 & = & r_1(t)+r_4(t)-(r_2(t)+r_3(t)) \ ,
\end{eqnarray}
where $X(s_i,t)$ is the concentration of protein $X$ in its machine state $s_i$ at time $t$. The parameter $k_{i}$ is the rate constant for an elementary reaction process defined by Rule $i$, and $r_i$ is the overall reaction rate for Rule $i$ (Table~\ref{tab:mod}).  One can systematically write down ODEs for other MFA states for the model of Fig.~\ref{fig:react}. We assume the total numbers of automata $A$ and $B$ are conserved, i.e., $A_{\rm tot}=A(s_1,t)+A(s_2,t)+A(s_3,t)$ and $B_{\rm tot}=B(s_1,t)+B(s_2,t)$. Note that $A(s_3,t)=B(s_2,t)$, the number of bonds formed between automata $A$ and $B$. Based on these constraints, only one more independent ODE is needed to describe the whole system:
\begin{eqnarray}
\frac{dA(s_3,t)}{dt} & = & k_3A(s_2,t)B(s_1,t)-k_4A(s_3,t) \notag \\ 
& = & r_3(t)-r_4(t) \ .
\end{eqnarray}
The above procedure of constructing a system of ODEs can be automated. In some systems modeled by MFAs, the construction of ODEs may not be straightforward. For example, if an MFA state transition depends on the status of a predicate evaluation, the calculation of the transition rate needs to be adjusted to account for the outcome of predicate evaluation. It is in many cases a difficult task to accurately account for rates of conditional transitions. We will discuss this issue below when we consider an example model for a MAPK cascade with a scaffold protein. 

\subsection{Stochastic simulation}

The temporal dynamics of a biochemical reaction system can be modeled as a continuous-time discrete-state Markov process to account for the evolution of the system configuration, which can be described by the following master equation~\cite{van2007stochastic}: 
\begin{equation}\label{eq:cke}
\frac{dp(c,t)}{dt}= \sum_{c'\ne c} \left[w(c|c')p(c',t) - p(c,t)w(c'|c) \right] \ ,
\end{equation}
where $p(c,t)$ is the probability that the system is found in configuration $c$, and $w(c'|c)$ gives the transition rate from configuration $c$ to $c'$. In a chemical reaction system, a configuration is defined by the concentrations of all chemical species. In a system specified by MFAs, a configuration $c$ is determined by the states and connectivities of MFAs. More precisely, the configuration is given by the properties of the individual MFA agents in the system, including the states and internal variables of these MFA agents. Analytical solutions to the above master equation are only possible for very simple systems. For a typical system, direct numerical integrations of the master equation is often intractable because of the enormous size of the configuration space. Kinetic Monte Carlo  simulation is applied to conduct sequential random walks through the configuration space and to obtain stochastic trajectories for a system of interest.

A system of rate processes can be simulated by the classic kinetic Monte Carlo method~\cite{voter2007introduction}. In our case, coupled processes in a biomolecular interaction system are defined by reaction rules that proceed in time at rates $\bf r$. These rule rates are determined by the current configuration of the system. At each time step, the waiting time $\tau$ for the next reaction event can be sampled from an exponential distribution with a mean waiting time of $1/r_{\rm tot}$, where $r_{\rm tot}=\sum_ir_i$ is the overall reaction rate of the system. To select a process that generates the next reaction event, one can sample a rule $i$ proportional to its rate $r_i$~\cite{yang2008kmc}.

For rule-based models defined by MFA structures, additional sampling steps are needed in each step to identify MFA agents that should undergo state transitions. Below, we outline a kinetic Monte Carlo algorithm for simulating systems specified in terms of MFAs.

\vspace{0.05in}

[Step 1] Initialization. Set time $t=0$, set the initial states and copy numbers of individual MFA agents, specify the rate constants of rate laws associated with reaction rules; calculate rule rates ${\bf r}$, and specify stopping criteria.

\vspace{0.05in}

[Step 2] At each time step, select a rule $i$ and a waiting time $\tau$, and update the time $t\leftarrow t+\tau$.

\vspace{0.05in}

[Step 3] Sample MFA agents from MFA candidates that are in permissible configurations as specified in the reaction rule sampled in Step 2, execute the state transitions of the sampled MFA agents, and recalculate the rate vector $\bf r$.

\vspace{0.05in}

[Step 4] Repeat Steps 2 and 3 until a stopping criterion is satisfied.

\vspace{0.05in}

In the above algorithm, Step 3 describes an agent-based simulation that tracks the states of individual proteins modeled by MFAs. A simulation produces single-protein configurations with details about reactive sites as well as connections between proteins.

Several general kinetic Monte Carlo methods for simulation of rule-based models have recently been developed~\cite{yang2008kmc,yang2008rfs,danos2007ssc,colvin2009simulation,colvin2010rulemonkey}, which can be readily adapted to suit the MFA framework. A rule-based kinetic Monte Carlo simulation involves sampling molecular agents or agent components that are permissible for transformation according to a rule~\cite{yang2008kmc}. In simulating an agent-based MFA model, after a rule is sampled in Step 2,  the algorithm described above involves searching for reactant agents in a population of candidate MFA agents that satisfy the rule protocol. The selected rule is executed by transforming the sampled MFA agents (as in Step 3 in the above procedure). MFA transformations are executed by sending input signals as specified in the rule to the sampled MFA agents. 

In some situations, the individual states of proteins and the connections of proteins within protein complexes may not be of interest, or such information may not be experimentally resolvable to verify the predictions generated by an agent-based simulation. In these cases, A kinetic Monte Carlo procedure that incorporates only observable quantities may be adopted. A kinetic Monte Carlo simulation can proceed as long as one is able to update the rates of reaction rules for each time step, which only requires tracking the populations of those machine states indicated in the rules (see Table~\ref{tab:mod}). These local MFA states usually consist of experimentally accessible quantities such as the number of protein bonds or phosphorylated sites. Other quantities, such as the concentration of a complex, may in some cases be synthesized from basic MFA configurations. 

\section{Example: MAPK cascade}
In this section, we use the example of a scaffold-mediated MAPK cascade to demonstrate how to construct and simulate an MFA-based model of a signaling pathway. 

The system is inspired by the scaffold-mediated MAPK cascade in yeast. The scaffold protein Ste5 possesses three domains that bind the MAP kinases Ste11 (MAPKKK), Ste7 (MAPKK) and Fus3 (MAPK) in the signaling pathway for the mating response~\cite{garrington1999organization}. We consider a scaffold protein with three independent binding sites: $\alpha$, $\beta$ and $\gamma$ sites, and three MAP kinases: a MAPKKK that binds to the $\alpha$ site of the scaffold protein, a MAPKK that binds to the $\beta$ site of the scaffold protein and can be phosphorylated by MAPKKK, and a MAPK that binds to the $\gamma$ site of the scaffold protein and can be phosphorylated by MAPKK. We assume that (1) binding reactions of the different kinases to the scaffold protein are independent processes, (2) MAPKKK can only be phosphorylated when it is bound to the scaffold protein, (3) phosphorylation of MAPKK (MAPK) can happen only when its kinase, phosphorylated MAPKKK (MAPKK), colocates on the same scaffold protein, and (4) phosphorylation and dephosphorylation of kinases can be modeled as single-step processes. 

\begin{figure}[tb]
\centering
\includegraphics[scale=0.5]{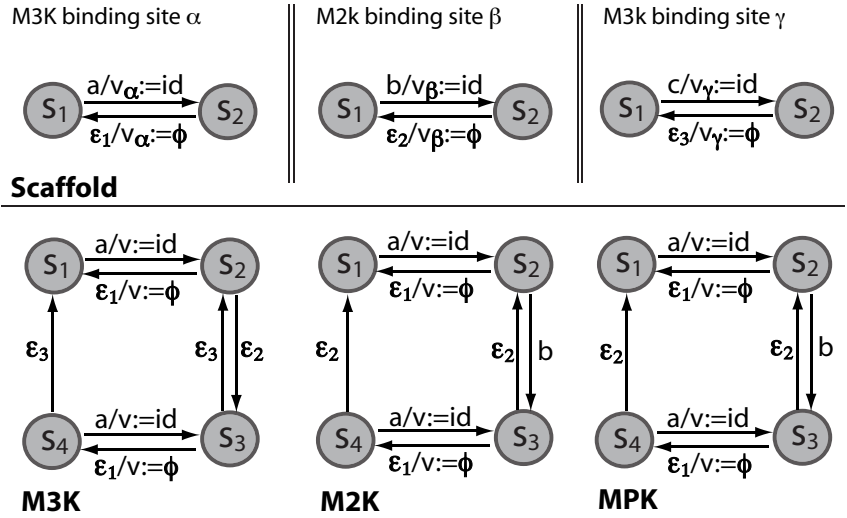}
\caption{\label{fig:mapk} MFA state transition diagrams for proteins in a MAPK cascade with a scaffold protein. The scaffold protein has three submolecular binding sites: the $\alpha$, $\beta$ and $\gamma$ sites that bind to M3K, M2K and MPK, respectively. State $s_1$ ($s_2$) for each scaffold site indicates that the site is free (bound). Each kinase has four states: $s_1$(free and unphosphorylated), $s_2$ (bound and unphosphorylated), $s_3$ (bound and phosphorylated), and $s_4$ (free and phosphorylated). Internal variables ($v_\alpha$, $v_\beta$, $v_\gamma$ in the scaffold, and $v$'s in the kinases) record contextual information, in particular, record binding partners. Note that the names of inputs and internal variables are local to the MFAs in which they appear. In other words, inputs or internal variables in different MFAs with the same name are not identical.}
\end{figure}

To simplify notations, we will use $\rm SCF$ to denote the scaffold protein and $\rm M3K$, $\rm M2K$ and $\rm MPK$ to denote MAPKKK, MAPKK and MAPK, respectively. Figure~\ref{fig:mapk} illustrates state transition diagrams of MFA models for the four proteins involved in the system. A total of 12 reaction rules describing protein interactions in the system are listed in Table~\ref{tab:mapk}.

In particular, we note that assumption (3) above requires checking molecular contexts to determine if protein state transitions are permissible, which exemplifies one important feature of MFA framework that allows modeling context-sensitive interactions. Consequently, Rules 9 and 11 in Table~\ref{tab:mapk} require predicate evaluations that examine nonlocal machine configurations. For example, Rule 9 validates an M2K agent by checking whether it is bound to an SCF agent that in turn is bound to an M3K agent in $s_3$. This contextual information is provided in the rule as a pattern of bond associations, M2K--SCF--M3K($s_3$). In general, a pattern of a multiprotein complex can be specified by a data structure that represents a connectivity graph.

\subsection{Deterministic ODEs}
\begin{figure}[t]
\centering
\includegraphics[width=6cm]{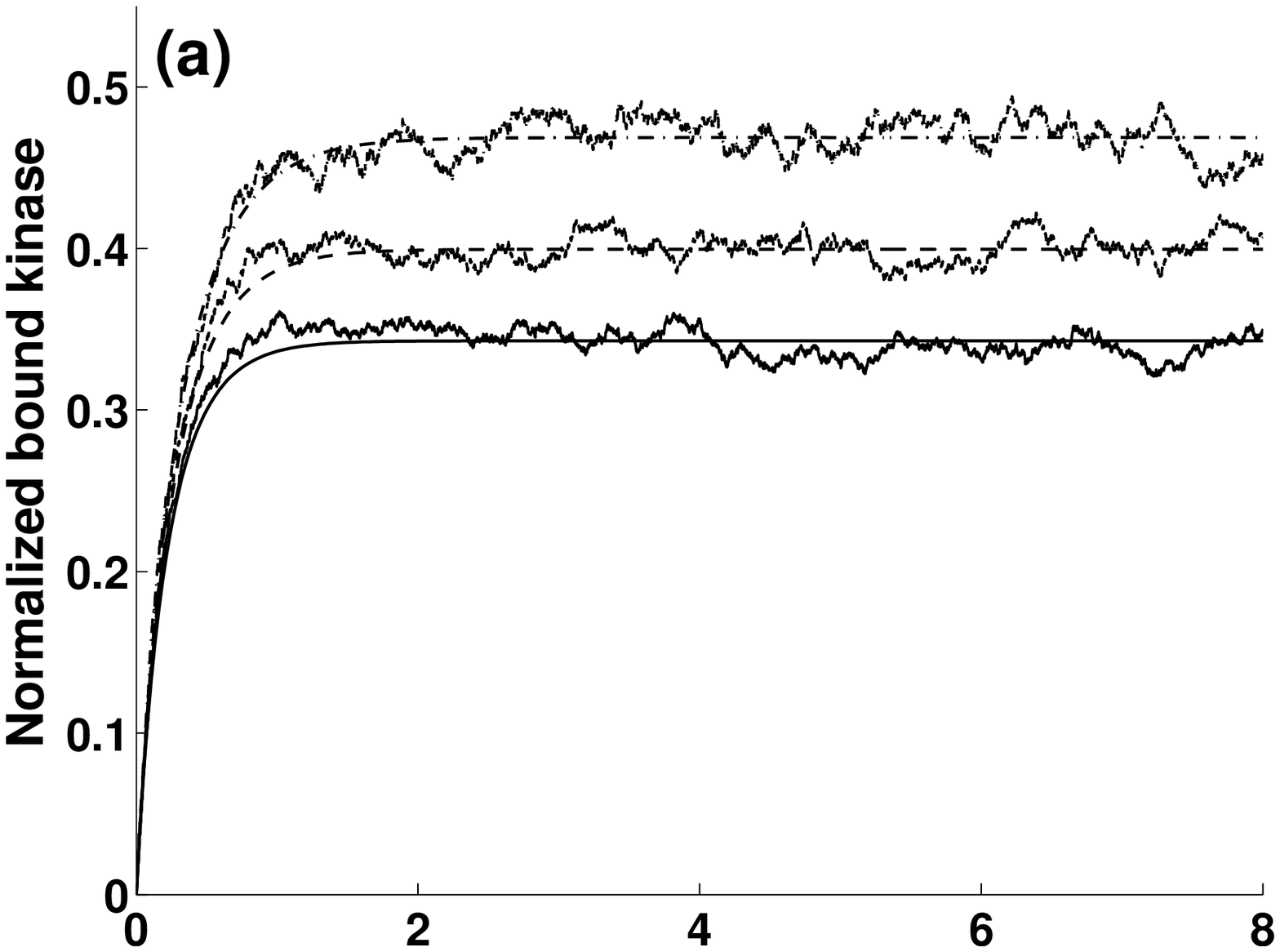}

\vspace{0.05in}

\includegraphics[width=6cm]{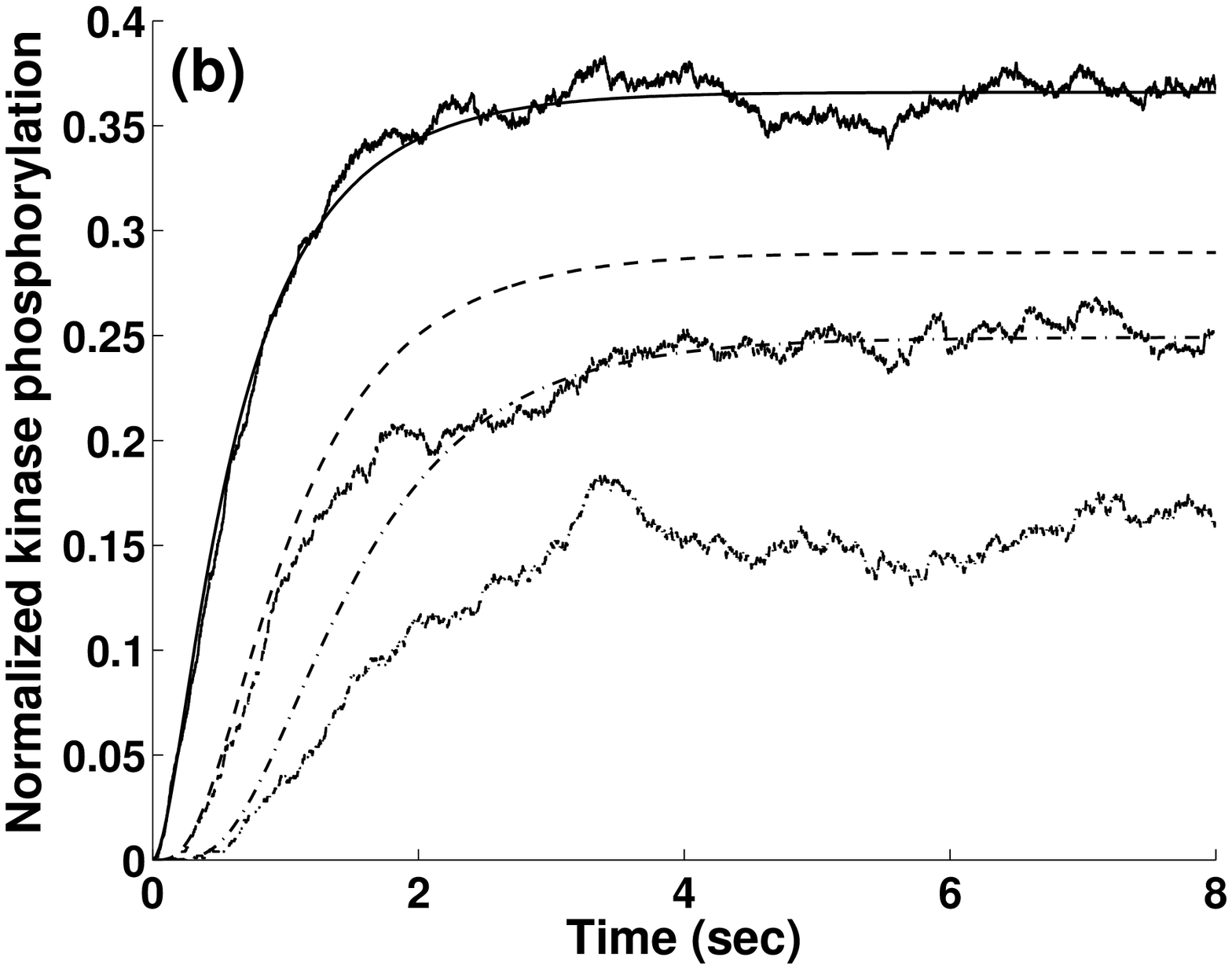}

\vspace{0.05in}
\caption{\label{fig:fksim} Simulation of the model for the MAP kinase cascade with a scaffold of Fig.~\ref{fig:mapk}. Results from ODE simulations (smooth curves) vs. those from kinetic Monte Carlo simulations (fluctuating curves). (a) Kinases bound to the scaffold ${\rm M3K}(s_2+s_3)$ (solid line), ${\rm M2K}(s_2+s_3)$ (dashed line) and ${\rm MPK}(s_2+s_3)$ (dash-dot) each normalized by total number of kinases. (b) Phosphorylated kinases ${\rm M3K}(s_3+s_4)$ (solid line), ${\rm M2K}(s_3+s_4)$ (dashed line) and ${\rm MPK}(s_3+s_4)$ (dash-dot) each normalized by total number of kinases. ${\rm SCF}_{\rm tot}=1000$, ${\rm M3K}_{\rm tot}=2000$, ${\rm M2K}_{\rm tot}=1500$ and ${\rm MPK}_{\rm tot}=1000$. Initially, all machines are in state $s_1$. Kinetic parameters ($k_1$ to $k_{12}$, defined in Table~\ref{tab:mapk}) are chosen for the purpose of illustration, not to model a specific system. $k_1=k_3=k_5=1.66\times 10^{-6}$ nL$\cdot$s$^{-1}$, $k_2=k_4=k_6=1.0$ s$^{-1}$, $k_7=k_9=k_{11}=3.0$ s$^{-1}$ and $k_8=k_{10}=k_{12}=1.0$ s$^{-1}$.}
\end{figure}

We first show how to write deterministic ODEs to describe the evolution of the population levels of MFA states. Our MFA model of the MAPK cascade consists of 15 independent machine states, compared to a total 33 distinct chemical species implied by the same set of reaction rules. Furthermore, we assume that the total amount of each protein is conserved, which corresponds to $M_{\rm tot}=\sum_{i=1}^nM(s_i)$ for an MFA with $n$ states, where $M_{\rm tot}$ is the total number of the MFA agents. Kinetic parameters that appear in the following equations are defined in Table~\ref{tab:mapk}. The following equation characterizes M3K binding to the $\alpha$ site of the scaffold protein:
\begin{equation}
\frac{d{\rm SCF}.\alpha(s_1)}{dt}=-k_1[{\rm M3K}(s_1)+{\rm M3K}(s_4)]{\rm SCF}.\alpha(s_1)+k_2{\rm SCF}.\alpha(s_2) \ .
\end{equation}
The equations for M2K and MPK binding to the $\beta$ and $\gamma$ sites of the scaffold are similar. Together with two algebraic relationships, ${\rm M3K}_{\rm tot}=\sum_{i=1}^4{\rm M3K}(s_i)$ and ${\rm SCF}.\alpha(s_2)={\rm M3K}(s_2)+{\rm M3K}(s_3)$, two additional ODEs are needed to completely account for the populations of the four possible states of machine M3K:
\begin{eqnarray}
\frac{d{\rm M3K}(s_1)}{dt} & = & k_2{\rm M3K}(s_2)+k_8{\rm M3K}(s_4)-k_1{\rm SCF}.\alpha(s_1){\rm M3K}(s_1)\\
\frac{d{\rm M3K}(s_2)}{dt} & = & k_1{\rm SCF}.\alpha(s_1){\rm M3K}(s_1)+k_8{\rm M3K}(s_3)-(k_2+k_7){\rm M3K}(s_2) \ .
\end{eqnarray}
For the case of M2K or MPK, because phosphorylation of a kinase bound to the scaffold (transition from state $s_2$ to $s_3$) is a conditional process that requires colocalization of an upstream kinase on the same scaffold protein, only a fraction of all kinases in state $s_2$ are candidates for transition to $s_3$. One can use the MFA state transition diagrams of Fig.~\ref{fig:mapk} to write the ODEs that track the populations of states $s_1$ and $s_2$ for M2K as follows:
\begin{eqnarray}
\frac{d{\rm M2K}(s_1)}{dt} & = & k_4{\rm M2K}(s_2)+k_{10}{\rm M2K}(s_4)-k_3{\rm SCF}.\beta(s_1){\rm M2K}(s_1)\\
\frac{d{\rm M2K}(s_2)}{dt} & = & k_3{\rm SCF}.\beta(s_1){\rm M2K}(s_1)+k_{10}{\rm M2K}(s_3)-(k_4+k_9\theta_1){\rm M2K}(s_2) \ .
\end{eqnarray}
Similar ODEs can be written to describe the dynamics of MPK states. The handling of Rule 9 (11) in Table~\ref{tab:mapk} for M2K (MPK) phosphorylation requires special attention. Not all M2Ks (or MPKs) in state $s_2$ are subject to transition to $s_3$ at a given time. The eligible fraction must satisfy the model assumption that requires colocalization of an M2K (MPK) with its enzyme, a phosphorylated M3K (M2K), on the same scaffold protein.  The factors $\theta_1$ and $\theta_2$ are introduced to account for the fractions of ${\rm M2K}(s_2)$ and ${\rm MPK}(s_2)$ that are eligible for transition to state $s_3$. In general, it is non-trivial to obtain analytical equations for factors such as $\theta_1$ and $\theta_2$. In this example, we simply approximate these factors as follows: $\theta_1\approx {\rm M3K}(s_3)/{\rm SCF}_{\rm tot}$ and $\theta_2\approx {\rm M2K}(s_3)/{\rm SCF}_{\rm tot}$. These approximations are exact only if kinase phosphorylation reactions are independent and context insensitive. Figure~\ref{fig:fksim} shows results from deterministic ODE simulations, compared to those from kinetic Monte Carlo simulations (performed as described below). We note that the stochastic simulation results are exact. The time trajectories for phosphorylated M3K from the deterministic and the stochastic simulations agree with each other on average. However, the ODE-based simulations of M2K and MPK phosphorylation deviate from those of the exact stochastic simulations due to the approximations used for $\theta_1$ and $\theta_2$ (Fig.~\ref{fig:fksim}(b)). Writing ODEs to directly describe MFA states as variables provides a way of quickly constructing a quantitative (sometimes approximate) model for simulation. One can avoid using approximations by expanding reaction rules into a chemical reaction network~\cite{faeder2005rule,blinov2004bionetgen}. In such cases, the variables in these ODEs would correspond to concentrations of chemical species rather than MFA states.  Generating these ODEs may be impractical, because the number of ODEs needed to capture the dynamics of the chemical species implied by a set of MFAs can be much larger than the number of ODEs needed to capture the dynamics of MFA states.

\begin{table*}[t]
{\centering \caption{\label{tab:mapk} Reaction rules for the MAPK cascade with a scaffold protein}
\footnotesize
\begin{tabular}{lll} \hline  {\bf Rule description} & {\bf Formal specification} & {\bf Rate law} \\ \hline\hline
1. M3K recruitment & \{$a,a$\}$\to$\{SCF.$\alpha,$M3K\}  & $k_1{\rm M3K}(\cdot){\rm SCF}.\alpha(\cdot)$\\
2. M3K dissociation & \{$\varepsilon_1,\varepsilon_1$\}$\to$\{SCF.$\alpha,$M3K\}$\backslash$SCF.$\alpha$--M3K & $k_2{\rm SCF}.\alpha(\cdot)$ \\
3. M2K recruitment & \{$b,a$\}$\to$\{SCF.$\beta,$M2K\} & $k_3{\rm M2K}(\cdot){\rm SCF}.\beta(\cdot)$ \\ 
4. M2K dissociation & \{$\varepsilon_2,\varepsilon_1$\}$\to$\{SCF.$\beta,$M2K\}$\backslash$SCF.$\beta$--M2K  & $k_4{\rm SCF}.\beta(\cdot)$ \\ 
5. MPK recruitment & \{$c,a$\}$\to$\{SCF.$\gamma,$MPK\} & $k_5{\rm MPK}(\cdot){\rm SCF}.\gamma(\cdot)$ \\
6. MPK dissociation & \{$\varepsilon_3,\varepsilon_1$\}$\to$\{SCF.$\gamma,$MPK\}$\backslash$SCF.$\gamma$--MPK & $k_6{\rm SCF}.\gamma(\cdot)$ \\
7. M3K phosphorylation & \{$\varepsilon_2$\}$\to$\{M3K\} & $k_7{\rm M3K}(\cdot)$ \\
8. M3K dephosphorylation & \{$\varepsilon_3$\}$\to$\{M3K\} & $k_8{\rm M3K}(\cdot)$ \\
9. M2K phosphorylation & \{$b$\}$\to$\{M2K$\}\backslash$M2K--SCF--M3K($s_3$) & $k_9{\rm M2K}(\cdot)$ \\
10. M2K dephosphorylation & \{$\varepsilon_2$\}$\to$\{M2K\} & $k_{10}{\rm M2K}(\cdot)$ \\
11. MPK phosphorylation & \{$b$\}$\to$\{MPK\}$\backslash$MPK--SCF--M2K($s_3$) & $k_{11}{\rm MPK}(\cdot)$\\ 
12. MPK dephosphorylation & \{$\varepsilon_2$\}$\to$\{MPK\} & $k_{12}{\rm MPK}(\cdot)$\\\hline
\end{tabular}}
\end{table*}

\subsection{Kinetic Monte Carlo simulation}

The stochastic simulations were carried out using the kinetic Monte Carlo algorithm described in the previous section. The system-specific implementation was an agent-based simulation procedure that samples the reaction rule list in Table~\ref{tab:mapk} and transforms individual MFA agents. We note that ${\rm M2K}_\theta (s_2)$ and ${\rm MPK}_\theta (s_2)$ in Rules 9 and 11 (Table~\ref{tab:mapk}), in contrast to the approximations used in the ODE simulations, are updated by on-the-fly bookkeeping that tracks reaction events immediately coupled to the two rules. The bookkeeping accounts for the numbers of M2K and MPK kinases in state $s_2$ that are eligible for the transitions specified by Rules 9 and 11. This implementation corresponds to a rejection-free algorithm for kinetic Monte Carlo simulation of rule-based models~\cite{yang2008rfs,colvin2010rulemonkey}, in which the exact rates of rules are calculated. This scheme becomes difficult to implement and computationally inefficient when predicates can be potentially affected by many types of reaction events. For example, Rule 9 may be affected by events from eight other distinct rules, as indicated in the dependence matrix $\mathcal{D}$ below. When the algorithm executes an event defined by any of these eight rules, the algorithm also needs to evaluate whether the predicate of Rule 9 is affected. 

Reaction rules are usually coupled in most systems. In other words, an event from one rule may affect the rates of others. Such coupling relationships between rules can be summarized by a ``dependency graph," or ``influence map"~\cite{danos2007rule}. For the example of the MAPK cascade model, we can summarize the rule dependence in the form of the following adjacency matrix:
\begin{equation}\label{eq:d}
\mathcal{D}=\left[\begin{array}{cccccccccccc}
1 & 1 & 0 & 0 & 0 & 0 & x & 0 & x & 0 & 0 & 0 \\
1 & 1 & 0 & 0 & 0 & 0 & x & 0 & x & 0 & 0 & 0 \\
0 & 0 & 1 & 1 & 0 & 0 & 0 & 0 & x & 0 & x & 0 \\
0 & 0 & 1 & 1 & 0 & 0 & 0 & 0 & x & 0 & x & 0 \\
0 & 0 & 0 & 0 & 1 & 1 & 0 & 0 & 0 & 0 & x & 0 \\
0 & 0 & 0 & 0 & 1 & 1 & 0 & 0 & 0 & 0 & x & 0 \\
0 & 0 & 0 & 0 & 0 & 0 & 1 & 1 & x & 0 & 0 & 0 \\
0 & 0 & 0 & 0 & 0 & 0 & x & 1 & x & 0 & 0 & 0 \\
0 & 0 & 0 & 0 & 0 & 0 & 0 & 0 & 1 & 1 & x & 0 \\
0 & 0 & 0 & 0 & 0 & 0 & 0 & 0 & x & 1 & x & 0 \\
0 & 0 & 0 & 0 & 0 & 0 & 0 & 0 & 0 & 0 & 1 & 1 \\
0 & 0 & 0 & 0 & 0 & 0 & 0 & 0 & 0 & 0 & x & 1
\end{array}\right] \ .
\end{equation}
A matrix entry $d_{ij}$ at row $i$ and column $j$ represents the influence of an event from rule $i$ on the rate of rule $j$. An entry with a boolean value 1 (0) indicates that an event in a rule has (does not have) an immediate impact on the rate of another rule, whereas $x$ indicates that an event may or may not alter the rate of another rule, depending on whether the event changes the predicate of the other rule. For the example of  the MAPK cascade, an event from Rule 1 can alter the rate of Rule 9 only in the case that a recruited M3K is phosphorylated and an unphosphorylated M2K is bound to the same scaffold agent. A dependency graph can in general be automatically derived by systematically analyzing a reaction rule set. In practice, the adjacency matrix $\mathcal{D}$ can be made more quantitative by replacing the Boolean value 1's with pre-calculated values of rate changes. For example, the entry $d_{89}$ (the influence of an event from Rule 8 on the rate of Rule 9) can be (conditionally) replaced by a numerical value, the value of $-k_9$ (the minus sign indicates a reduction in the rule rate), because an M3K dephosphorylation may decrement the eligible population of M2K agents.

To reduce bookkeeping, one may use a rejection algorithm~\cite{yang2008kmc} as an alternative. This algorithm allows one to use rejection sampling to simplify the firing of reactions when state transitions are associated with predicate functions. In this algorithm, one calculates rule rates and samples participant MFA agents without considering constraints imposed by the predicate functions. Sampled trial agents are rejected for state transition if the predicate function does not evaluate as true. For example, the number of all MPK agents in state $s_2$ can be used to calculate the rate of Rule 11 as $\widetilde{r}_{11}=k_{11}{\rm MPK}(s_2)$. Once an event from Rule 11 is sampled, a trial MPK agent in state $s_2$ will be chosen, which will then undergo a state transition only if the transition condition is satisfied. Otherwise, the transition is rejected. Implementation of a rejection algorithm is easier compared to that of a rejection-free algorithm. The rejection algorithm enforces the conditions of state transitions only when a rule that has conditional transitions is sampled. Our experience suggests that a rejection algorithm is more computationally efficient than a rejection-free algorithm as long as rejections do not comprise the vast majority of all Monte Carlo steps~\cite{yang2008rfs}.

The sparsity of dependency matrix $\mathcal{D}$ in Eq.~(\ref{eq:d}) (with many entries being zeros) indicates that rule couplings are largely localized. Similar to an efficient implementation of a conventional stochastic simulation algorithm for chemical reaction systems~\cite{gibson2000efficient}, for a large system specified by a considerable number of rules, weak dependency between rules can be used to optimize the procedure for updating rule rates after each Monte Carlo step.

\section{Discussion} 

Information processing in living cells can be viewed as protein computation, in which proteins act as computing machines that react to external signals by local computation~\cite{regev2002cac,bray1995protein}. Thus, a protein-protein interaction system is an integrative and distributed system with numerous computing devices interacting with each other under certain protocols. This perspective suggests that formal computing models can help archive, organize and interpret protein functions and their interactions. Formal structures also facilitate the process of computational modeling of complex and large-scale protein-protein interaction systems. 

In this work, we introduce an extension to the traditional computing model of finite state automata to describe protein behaviors in response to external inputs and protein interactions. The MFA formalism offers a rule-based platform for modeling and simulating biochemical systems, especially for signal transduction. An MFA is in essence a data structure that specifies protocols to define the activity of a protein in a discrete state space. An MFA can be used as a representation of knowledge or hypotheses about a protein and can serve as a building block for biomolecular interaction models. At the systems level, reaction rules connect separate MFA-represented proteins and model the dynamics of a protein interaction system as synchronized MFA state transitions. The MFA formalism allows for a clearer and more natural representation of proteins and the combinatorics of protein interactions. The MFA formalism adds in intuitive formulation of mechanistic models of signal transduction, which can be accessible for those who are familiar with the biological knowledge underlying the models. For quantitative model computation, as our example model of a MAP kinase cascade system demonstrates, ODEs for deterministic simulations can be constructed to track concentrations of machine states that often directly correspond to experimentally resolvable quantities. (For some states, the ODE solutions give approximations.) For exact and stochastic simulations, rule-based kinetic Monte Carlo methods can be applied.

Formalisms derived from finite automata theory have been proposed earlier for applications in biology. Notably, Harel and coworkers~\cite{fisher2009} have applied statecharts, a graphical and hierarchical (extended) finite automata structure [50], to model and simulate cell development and dynamics of cell populations. Recently, use of finite automata to model biomolecular interactions was studied by Cardelli~\cite{cardelli-artificial}, in which the author introduced the concept of Òpolyautomata.Ó The concepts of polyautomata and MFA are closely related. In Ref.~\cite{cardelli-artificial}, polyautomata are used to represent SPiM scripts, which specify stochastic simulations via the stochastic pi-calculus approach implemented in the SPiM software tool~\cite{phillips2007efficient}. Here, we have shown that MFAs can be used to specify both deterministic and stochastic simulation approaches (Fig. 6).  Cardelli~\cite{cardelli-artificial} demonstrated that polyautomata are useful for modeling protein complex formation, including polymerization-like reactions.  Here, we have shown that MFAs can be used to model protein complex formation as well as post-translational modifications of proteins, as illustrated in Figs. 3 and 5.  Finally, the MFA formalism extends the polyautomata concept in an important way by allowing for the explicit representation of the functional components of proteins and the structural relationships among these components (Fig. 2).

A potential strength of the MFA framework is the mature development and many applications of finite automata. Various forms of finite state automata have been industry standards for modeling reactive systems for many years.  In particular, MFAs are amenable to hardware implementations using programmable logic devices including widely-used programmable array logic (PAL), generic array logic (GAL) and field programmable gate array (FPGA) devices. In particular, because of the sequential nature of discrete event-driven simulation, the performance of simulating large-scale complex biochemical reaction systems by stochastic simulation is poor even if it is not prohibitive. Hardware implementation may yield an advantage in speed. The first hardware (FPGA) stochastic simulations of biochemical networks were implemented by Keane and co-workers~\cite{keane2004compiled} and Salwinski and Eisenberg~\cite{salwinski2004silico}. Taking advantage of the parallel architecture of FPGA, implementations by Salwinski and Eisenberg~\cite{salwinski2004silico} allowed improvement in the speed of simulation of a simple bimolecular reaction by at least one order of magnitude compared to a conventional software implementation on a benchmarking platform. Recently, Yoshimi et al.~\cite{yoshimi2007fpga} implemented the next reaction method of Gibson and Bruck~\cite{gibson2000efficient} in an FPGA-based simulator and were able to achieve a significant speed-up. Implementation of rule-based models into programmable circuits has yet to be realized, but finite state machines are routinely implemented in hardware. Hardware implementation stores state variables and embeds state transition protocols into digital electronic circuits. In the design process, the MFA structures need to be translated into binary logic to apply digital computing to achieve the defined machine dynamics. As proteins can be modeled as MFAs, in principle, the dynamics of a protein interaction network may be simulated by electronic circuits.

Another potential use of the MFA structure is to archive proteins in terms of their reaction dynamics. Since an MFA is a standalone structure for storing the discrete dynamics of a protein, we expect that it can be used to systematically archive protein functions, with the MFA structure serving as an elementary record type for a database. Protein records in current protein databases are mostly annotations including amino acid sequences, functional domains, associated functions, etc. However, such information cannot be readily used to construct mechanistic biomolecular interaction models. The MFA structure offers an alternative way of storing protein records. Using a database with MFA records, one can construct a model by querying the database for MFA structures to obtain a set of desired molecular building blocks. One can then specify reaction rules to connect these machines. Such a rule-based model can be efficiently revised and incrementally refined when records of MFAs involved in the model are updated to reflect new knowledge.

\section*{Acknowledgment}
This work was supported by grants GM076570, GM085273 and RR18754 from the National Institutes of Health, by the Department of Energy through contract DE-AC52-06NA25396, and by grant 30870477 from the National Science Foundation of China to J.Y. We thank the Center for Nonlinear Studies for support that made it possible for J.Y. to visit Los Alamos.

\end{document}